\documentclass[12pt,preprint]{aastex}

\def\psr{PSR~B0540$-$69}

\shortauthors{Johnston \& Romani}
\shorttitle{Giant Pulses from \psr{}}

\begin{document}
\title{Giant Pulses from \psr{} in the Large Magellanic Cloud}

\author
{Simon~Johnston\altaffilmark{1} and  Roger~W.~Romani\altaffilmark{2}}
\altaffiltext{1}{School of Physics, University of Sydney, NSW 2006, Australia.
email simonj@physics.usyd.edu.au}
\altaffiltext{2}{Dept. of Physics, Stanford University, Stanford,
CA  94305-4060, USA. email rwr@astro.stanford.edu}
%
%
%
%
%
%

\begin{abstract}
We report the discovery of the first giant pulses from an extragalactic
radio pulsar.
Observations of \psr{} in the Large Magellanic Cloud
made with the Parkes radio telescope at 1.38 GHz show single pulses
with energy more than 5000 times that of the
average pulse energy. This is only the second young pulsar, after the Crab,
to show giant pulse emission. Similar to the Crab pulsar, the giant pulses
occur in two distinct phase ranges and have significant arrival 
time jitter within these
ranges. The location of the giant pulses appears to lag the
peak of the (sinusoidal) X-ray profile by 0.37 and 0.64 phase,
although absolute timing between the radio and X-ray data is not yet secure.
The dispersion measure of the giant pulses
is 146.5~cm$^{-3}$pc, in agreement with the detection of the pulsar
at 0.64~GHz by Manchester et al. (1993). The giant pulses are
scatter broadened at 1.4 GHz with an exponential scattering time
of 0.4~ms and have an emission bandwidth of at least 256 MHz.
In 8~hr of integration we have failed to detect any integrated flux density
from the pulsar to a level of 13~$\mu$Jy, assuming a duty cycle
of 10\%. This implies the spectral
index between 0.64 and 1.38~GHz is steeper than --4.4.
\end{abstract}

\keywords{neutron stars: individual: \psr{}}

\section{Introduction}
\psr{} was discovered in the early 1980s by Seward, Harnden \& Helfand
(1984) using data from the Einstein X-ray Observatory.
The pulsar is located inside the supernova remnant SNR 0540--693
in the Large Magellanic Cloud. It has a short rotation period ($\sim$50~ms)
and a rapid spindown with a characteristic age of only 1500~yr.
There are strong similarities between this system and the Crab.
The pulsars have similar characteristics and the plerion around
\psr{} is very similar to the Crab Nebula in size and energetics
(Manchester, Staveley-Smith \& Kesteven 1993).
Following the discovery of the X-ray pulsations from \psr{},
optical pulsations were soon detected
(Middleditch \& Pennypacker 1985),
but the radio pulsations remained elusive. Finally,
Manchester et al. (1993; hereafter M93) discovered a
broad, weak radio pulse at 0.64 GHz. The flux density 
at that frequency is 0.4~mJy
and the pulse duty cycle is more than 80\% with the hint of a 
double profile. This is similar to the high energy profiles 
(de Plaa, Kuiper \& Hermsen 2003)
which also appear to show a broad, unresolved double structure.
M93 failed to detect the pulsar at either 1.4 or 0.44 GHz.

In comparison, the radio profile of the Crab pulsar is complex. It consists of
a so-called pre-cursor component and main and interpulse components.
The pre-cursor
has a very steep spectral index rendering it undetectable above
about 1 GHz. The main and interpulse emission are in phase with
the X-ray and $\gamma$-ray emission which both have narrow peaks.
At frequencies above 5 GHz, further components have been detected
(Moffett \& Hankins 1999).
The Crab pulsar also produces
giant pulses when the emission can exceed the mean pulse
energy by a factor of 1000 or more.
This giant pulse emission occurs at the phases of the main pulse and
interpulse but not at those of the pre-cursor or the `anomalous' high
frequency components
(Lundgren et al. 1995; Hankins et al. 2003).

Giant pulses have now also been discovered from the millisecond
pulsars PSRs B1937+21 (Cognard et al. 1996) and B1821--24
(Romani \& Johnston 2001). Cognard et al. (1996) noticed
that when pulsars are ordered by their magnetic field strength at
the light cylinder, the pulsars with giant pulses are highly ranked.
Romani \& Johnston (2001) noted the similarity between the phases of
the giant pulse emission and the X-ray emission in these pulsars and
speculated that the giant pulse emission may also originate from near the light
cylinder. McLaughlin \& Cordes (2003) have very recently reported the
non-detection of giant pulses from \psr{} through observations made at
0.66 and 1.52 GHz.

\section{Observations and Data Reduction}
Observations were made on 20 May 2001 (MJD 52049), with the Parkes 64-m
radio telescope. We used the center beam of the 21-cm multi-beam system
at an observing frequency of 1.38~GHz. The receiver had a system equivalent
flux density of 27~Jy on cold sky. The back-end consisted of a filterbank
system containing 512 channels per polarization, each of width 0.5 MHz for
a total bandwidth of 256~MHz.  The polarization pairs were summed to
form the total power, each
output was then sampled at 80 $\mu$s, one-bit digitized,
and written to DLT for off-line analysis. A total of 8~hr of
integration on \psr{} was obtained, split into 4 individual observations,
each of duration 2~hr.

To search for giant pulses, it is not necessary to know the rotational
frequency of the pulsar but it is necessary to know the
dispersion measure (DM). As an initial trial, we used the DM of 146~cm$^{-3}$pc
given in M93. After applying the appropriate delay to each frequency
channel, the 512 channels were summed.
The mean and rms of groups of 8192 samples (0.65~s) were examined
and those which showed obvious signs of interference were discarded.
Our nominal 5$\sigma$ sensitivity in
80~$\mu$s is 0.9~Jy. In practice, even after the removal of
high sigma points (clipping), we experienced
substantially larger background fluctuations and the rms exceeded our
expected rms by a factor of $\sim$2. The entire 8 hr observation
(3.6$\times 10^{8}$ samples) was then searched for peaks.
This technique was successfully used to discover giant pulses in
PSR~B1821--24 (Romani \& Johnston 2001) and the giant 
micro-pulses in PSR~B1706--44 (Johnston \& Romani 2002).
Two large amplitude, resolved pulses were discovered.
We were able to determine the DM of these single pulses by
subdividing the total bandwidth into 16 and fitting a straight line
to their time-of-arrival across the 16 sub-bands.
The DM of both pulses was determined to be 146.5$\pm$0.2~cm$^{-3}$pc.
Armed with this knowledge we then applied a DM of 146.5~cm$^{-3}$pc
to the entire data set.

We then used a more sophisticated search technique to detect weaker
giant pulses. The DM introduces a broadening of 220~$\mu$s across
each 0.5~MHz channel and
the full width at half maximum of the strong giant pulses 
was $\sim$600~$\mu$s, or 8 samples.
We therefore first smoothed the data using a top hat function with a
width of 8 samples and then searched the data looking for 4 consecutive
data points more than 3$\sigma$ from the mean. 

\section{Pulse Timing and Phase Comparisons}
In order to obtain the correct pulsar phase for each time sample, an accurate
timing solution for the pulsar is required. The latest published
ephemeris for \psr{} from X-ray timing with the Rossi X-ray Timing Explorer
(RXTE) does not cover the epoch of our observations (Zhang et al. 2001)
and we therefore used an updated ephemeris
provided to us by F.~Marshall and W.~Zhang from RXTE observations near
MJD 52050, including overlap with our observing epochs.  The peak of
the X-ray profile at the solar system barycenter occurred
0.0366~s after MJD 52050 and at this epoch the spin frequency was
19.7886114394~s$^{-1}$  and the frequency derivative
was $-$1.875580$\times 10^{-10}$~s$^{-2}$.
The position of the pulsar used in the ephemeris was
that obtained from optical observations by Caraveo et al. (1992).
Sample RXTE arrival times on the same dates as the radio observations
agree with the adopted ephemeris, with uncertainties $\la 0.5$ms.
Using this timing solution we were therefore able to
form an integrated profile of the entire 8~hr observation, and we assert
that uncertainties in the X-ray ephemeris would be insufficient to
broaden the pulse by more than $\sim$0.5~ms ($\sim$1\% of the
pulse period).

Direct comparison of the phases of the radio giant pulses with the X-ray
profile is, however, subject to some uncertainty. We believe the absolute
timing in the radio to be correct, as we have obtained arrival times
for both the Crab pulsar and PSR B1937+21 which agree with 
independent radio ephemerides. However, the absolute X-ray timing is 
not secure and it is known that there
are problems comparing RXTE timing of \psr{} and those made with the
Chandra X-ray Observatory (A. Rots, private communication).

\section{Integrated Profile}
We failed to detect any significant integrated pulsed emission from the pulsar.
The bottom panel of Fig.~1 shows the pulse profile with 
128 phase bins after 8~hr
of observations. The 1-$\sigma$ rms is 80~$\mu$Jy.
As a check against the possibility that the X-ray ephemeris was incorrect
we also performed a Fast Fourier Transform on each de-dispersed 2~hr dataset
independently. No significant signals were detected near the
expected pulsar frequency.
Assuming the pulsar has a duty cycle of $\sim$10\%, the upper limit
on the continuum flux density at 1.38~GHz is therefore only 13~$\mu$Jy compared
to 400~$\mu$Jy at 0.64~GHz (M93). This implies a spectral index 
steeper than --4.4 between the two frequencies, although this could be
relaxed if the duty cycle of the profile is large and/or the 0.64 GHz
flux density is incorrect.
This is extremely high, even by
pulsar standards, but it is comparable with the spectral index
of the pre-cursor component in the Crab pulsar.
The distance to the Large Magellanic Cloud is 49.4~kpc, hence
the luminosity of the pulsar at 1.38~GHz is less than 32~mJy~kpc$^{2}$.
Although this is low compared to the luminosity of other young
pulsars, and a factor of 100 less than the luminosity of the Crab
pulsar, it is significantly higher than that of the recently
discovered PSR~J0205+6449 in the SNR 3C58
(Camilo et al. 2002).

\section{Giant Pulses}
As described above, the initial search of the data detected 
two giant pulses. These are shown in the top panel of Fig.~1 where 
both the entire profile is shown and a more detailed view of 
the structure in the giant pulses.
The widths of these pulses are entirely consistent with an 
intrinisic delta function convolved with the sampling
rate (80~$\mu$s), the dispersion delay (220~$\mu$s) and
an exponential tail with a 1/e time of 0.4~ms. This exponential
broadening is likely the result of interstellar scattering.
The two pulses are offset in phase by 0.03.

Using the analysis described in Section 2 (data smoothing plus
searching for several large consecutive samples), we searched a range of 
DMs in order to determine the cut-off above which no spurious signals
were detected. The best results were obtained using the criteria that four
consecutive samples should exceed 3$\sigma$.
We applied this analysis on the data at
a DM of 146.5~cm$^{-3}$pc and detected 15 giant pulses down to this
sensitivity limit. 
Fig.~2 shows the flux density as a function of phase for
the 15 detected giants superposed on the X-ray profile taken from
De Plaa et al. (2003).
There are two separate phase regions for the giant pulses,
the first between phases 0.36 and 0.45 and the second between
0.61 and 0.66 phase.
These phases should be considered preliminary in light of the discussion
about absolute timing above.
This is similar to the Crab giant pulses
which also occur in two separate phase regions. However, the jitter
in the phases of the giants in \psr{} is much larger than 
in the Crab; $\sim$5~ms compared to $\sim$0.5~ms.
Fig.~3 shows the histogram of the flux densities obtained. Although we
are dealing only with small number statistics, it seems clear that
the slope is not as steep as the slope of --3.5 seen in the Crab pulsar
(Lundgren et al. 1995) or that of --2.8 seen in
PSR B1937+21 (Kinkhabwala \& Thorsett 2000).

Two other points are worth noting. The first is that the brightest
giant obtained has a flux density which exceeds the mean single
pulse flux density by a factor $\ga$5000. This is a larger factor
than any giant pulse seen in the Crab pulsar. All the giants detected
here exceed
the mean flux density by 700. Expressed in these terms, this rate
of giants (1 in 35000) exceeds that of the Crab pulsar (1 in
350000; see Lundgren et al. 1995) by an order of magnitude.
However, in terms of the emitted energy, the giants in \psr{} are
significantly less powerful than in the Crab pulsar. 

Finally, the non-detection of giant pulses in \psr{} at 1.5 GHz by
McLaughlin \& Cordes (2003) can probably be attributed to their
observational setup. Their sampling rate was 0.6~ms, however, the dispersion
smearing across each of their 5~MHz filters was $\sim$2~ms and this
is sufficient to dilute the giant pulse enough for it to fall below their
quoted detection threshold of 0.4~Jy. Their non-detection of
giant pulses at 0.66~GHz would
imply that either the spectral index of the giant pulses are shallower
than --3, in contrast to the integrated profile, or that there is no
giant pulse emission at this frequency. The latter seems unlikely given that
both the Crab and PSR~B1937+21 show giant pulse emission at frequencies
well below 0.66~GHz. McLaughlin \& Cordes (2003)
do not comment on a detection or non-detection of the integrated pulse 
at this frequency.

\section{Discussion}
\psr{} is the 4th pulsar known to emit giant pulses and there are
now two young pulsars and two millisecond pulsars which emit giants.
We detected the giant pulses {\it without} detecting the underlying
integrated pulsed emission.
The four pulsars are in the top five pulsars when ranked by
magnetic field at the light cylinder; only PSR~B1957+20 from this list
does not show giant emission. Clearly therefore this parameter is
important to our understanding of the giant pulse phenomenon.
Furthermore, all four objects have pulsed, non-thermal, X-ray emission.
In the Crab, PSRs~B1937+21 and B1821--24 the X-ray emission appears
to be in phase with the giant radio emission.
It is intriguing that for \psr{},
the two giant pulse windows are $\sim 0.5$ out of phase with the X-ray pulse
and that their spacing is similar to that of the high energy pulse components.
Whether this indicates an error in the relative phasing between the
radio and X-ray obserations or an association
of the giant pulses with the magnetic pole opposite the high energy emission
remains to be tested with independent phase comparisons.

Given the exponential broadening of the giant pulses at 1.38~GHz,
it seems likely that the integrated profile at 0.64~GHz
is dominated by scattering (M93). Scatter broadening scales with the inverse
fourth power of the frequency and hence should be $\sim$20 times larger
at 0.64~GHz than at 1.38~GHz. The 0.64~GHz profile is thus consistent with
a double peaked profile separated by $\sim$0.3 phase and 
broadened by $\sim$10~ms of scattering. It is tempting to relate the
location of this double peaked structure with the location of the
giant pulses detected at the higher frequency. Further observations at
0.64~GHz are needed to confirm this.
Furthermore, the scattering at high frequency would also explain the 
non-detection of the pulsar at 0.44~GHz reported by M93, where the 
scattering time would exceed the pulse period. These values also
indicate that the effects of scintillation are not important in the
giant pulse search toward \psr{}, contrary to the suggestion of
McLaughlin \& Cordes (2003).

The DM is high compared to the other known
pulsars in the Large Magellanic Cloud which have DMs in the range
67 -- 107~cm$^{-3}$pc (McConnell et al. 1991).
A scattering time of $\sim$0.4~ms at 1.38~GHz is also large, especially
when the scattering screen is likely to be located much closer to
the pulsar (i.e. in the LMC) than to the Earth.
The situation is analogous to that
seen in the Vela pulsar, where both the DM and the scattering are
much larger than expected for its distance. Both the excess DM and
the scattering likely arise inside the Vela SNR itself where the line
of sight intersects a dense clump of ionised material.

Expressed as a continuum flux density, the largest giant pulses have 
1.4 GHz flux densities at a distance of 1 kpc
of 4000, 200, 7 and 10 Jy for the Crab, PSRs \psr{}, 1821--24 and 1937+21
respectively for roughly the same number of rotations of the pulsar
($\sim$10$^{6}$).
The giant pulses from the two millisecond pulsars have 
very similar flux densities and these are
significantly less than those of the young pulsars.
From the point of view of searching for giant pulses, the peak flux
is the important value. 
The giant pulses in the Crab pulsar and in PSR~B1937+21 are
known to have widths of order microseconds. Assuming all the giant
pulses have durations of $\sim$1~$\mu$s, then the giant pulses have
peak fluxes at 1.4 GHz at a distance of 1 kpc of
130, 11, 0.03 and 0.02 MJy for the Crab, PSRs \psr{}, 1821--24 and 1937+21
which makes the differences between the young pulsars and the
millisecond pulsars even more striking.
In principle, therefore, the Parkes telescope is capable of detecting
giant pulses from young pulsars at distances of $\sim$1~Mpc, and
millisecond pulsars out to $\sim$50~kpc. In practise, however, lack
of knowledge of the DM and scatter broadening both serve to weaken
these limits. The prospects for future generation radio telescopes
are excellent however. Even if, in young pulsars, the giant pulses
only occur over the first $\sim$2000~yr, one might expect to see
$\sim$20 per large galaxy and more in the nearby starburst galaxies.
For millisecond pulsars, the faster objects which have high magnetic fields 
at the light cylinder, will produce giant pulses for a Hubble time.
There therefore may be a large population of millisecond pulsars which
produce giant pulses, enhancing the prospects for targetted searches
of globular cluster in external galaxies.
Detection of extra-galactic pulsars will yield the electron density
content of the inter-galactic medium, important in understanding
the nature of the dark matter (Maloney \& Bland-Hawthorn 2001).

\section{Summary}
We have detected giant pulses from the young pulsar \psr{}
in the LMC and yet have failed to detect the integrated flux 
from the pulsar after 8~hr observing. Four pulsars, including
\psr{}, are now known to
have giant pulses and these four are in the top five when ranked by
their magnetic field at the light cylinder.
In the near future we plan to observe \psr{} at both lower
and higher radio frequencies to determine the spectral index of
the giant pulse emission and also to replicate the detection
of the pulsar at 0.64~GHz by M93. We also plan a simultaneous
radio and X-ray observation to both obtain accurate phasing between
the radio and X-ray and to look for giant pulses in the X-ray.

\acknowledgements
The Australia Telescope is funded by the Commonwealth of 
Australia for operation as a National Facility managed by the CSIRO.
We thank F.~Marshall for providing sample RXTE pulse arrival times.
We thank A.~Lyne and D.~Nice for providing timing ephemerides for
the Crab and PSR~B1937+21.

\clearpage
\newpage
\begin{figure}
\plotone{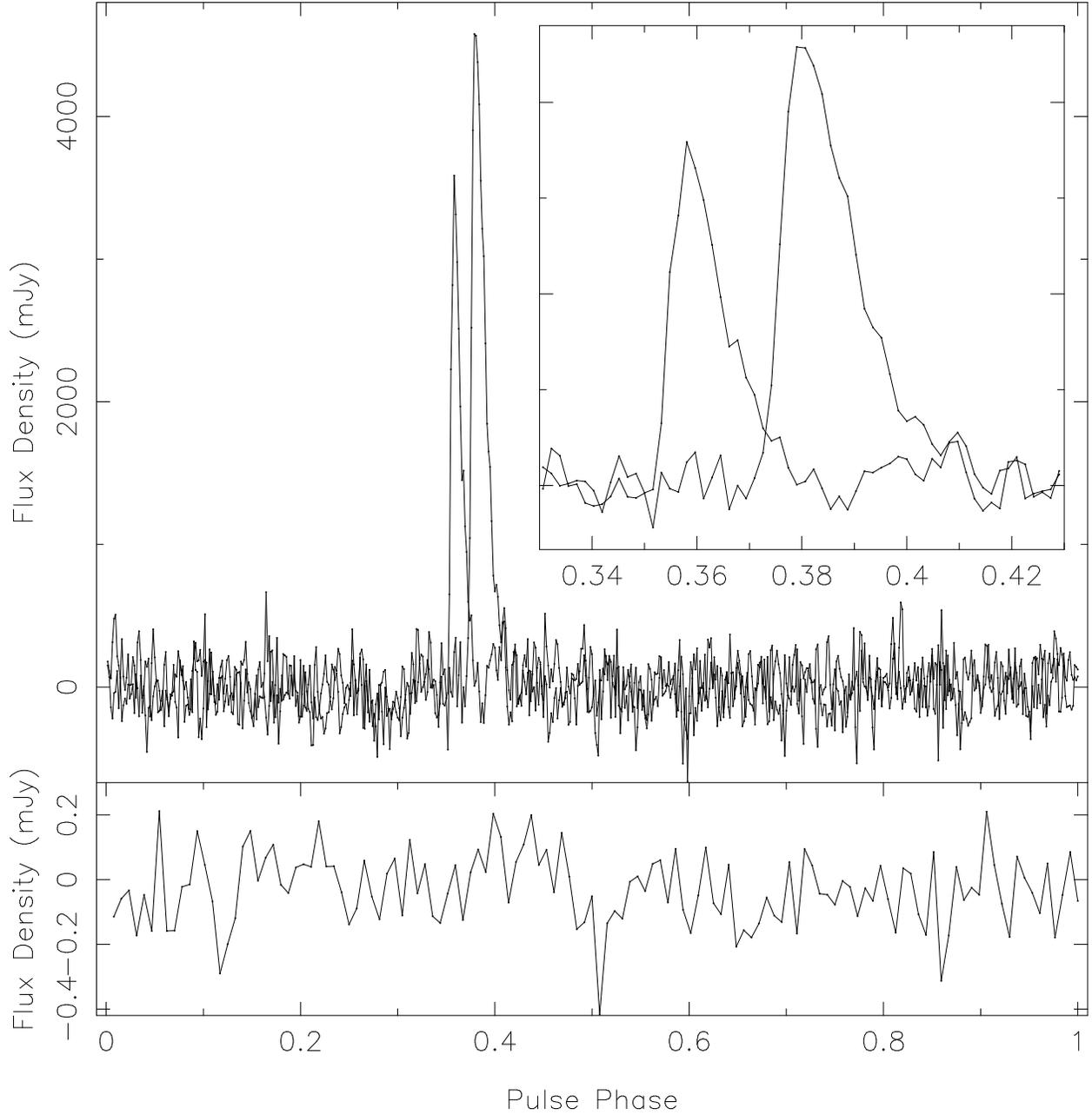}
\caption{The bottom panel shows the profile of \psr{} at 1.4 GHz 
after 8 hr of integration. The full pulse period is shown subdivided 
into 128 phase bins. The flux density limit on pulsed emission with a 10\%
duty cycle is 13~$\mu$Jy. The top panel shows
the two brightest giant pulses from \psr{} at the full resolution of
80~$\mu$s. The insert panel, with a restricted
phase range, shows the sharp rise and exponential decay of the pulses.
The peak of the X-ray emission occurs at phase 0.0 according to the
ephermeris given in Section 2 although we caution there may 
be an unknown (significant) offset between the X-ray and radio timing.}
\end{figure}
\clearpage
\newpage
\begin{figure}
\plotone{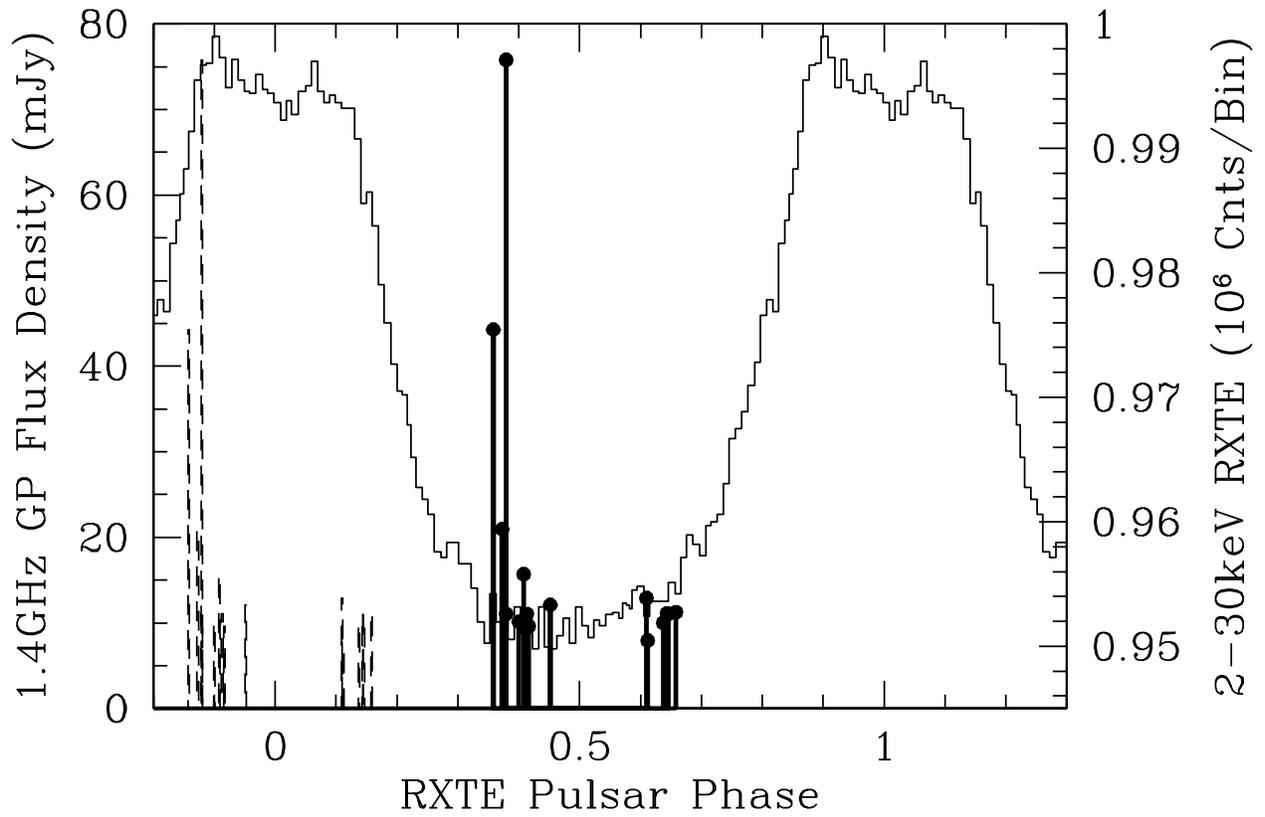}
\caption{RXTE X-ray light curve (1.5 periods, after de Plaa
{\it et al} 2003) with the nominal relative phases of the giant pulses shown
as bold lines.  Dashed lines give the giant pulses shifted in phase by 0.5,
showing the similarity between the spacing of the giant pulse phase windows
and the peak structure in the high energy light curve.}
\end{figure}
\clearpage
\newpage
\begin{figure}
\plotone{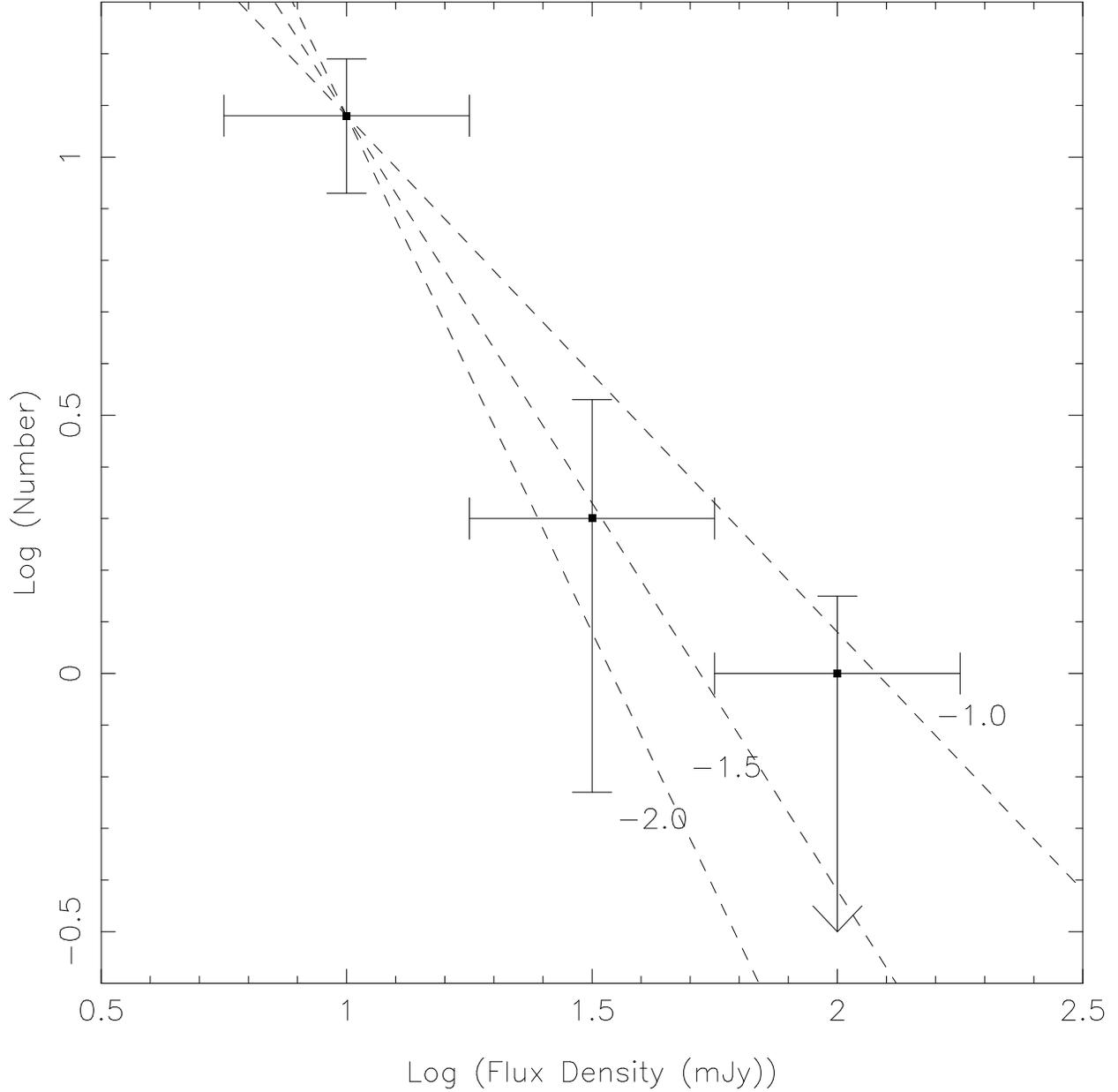}
\caption{Log N versus Log S for the 15 giant pulses along with sample slopes.
The error bars in Log N simply reflect $\sqrt{N}$ statistics and
those in Log S reflect the flux ranges used in the binning.}
\end{figure}
\end{document}